\documentclass[fleqn,10pt]{wlscirep}
\usepackage[utf8]{inputenc}
\usepackage[T1]{fontenc}

\usepackage{float}
\usepackage{graphicx}
\usepackage{bm}
\usepackage{xcolor}
\usepackage{tikz}
\usepackage{mathtools}
\usepackage{multirow}
\usepackage{array}
\usepackage{caption}
\usepackage{subcaption}
\usepackage{enumitem}
\usepackage[normalem]{ulem} 

\renewcommand{\thesection}{\Roman{section}} 
\newcommand{\lroman}[1]{\romannumeral #1}

\title{Prepare-and-measure and entanglement simulation beyond qubits}

\author[1,*]{Mani Zartab}
\author[1,2]{Giulio Gasbarri}
\author[1,3]{Gael Sentís}
\author[1]{Ramon Muñoz-Tapia}
\affil[1]{Física Teòrica: Informació i Fenòmens Quàntics, Departament de Física,
	Universitat Autònoma de Barcelona, Bellaterra (Barcelona), 08193, Spain}
\affil[2]{Naturwissenschaftlich-Technische Fakult{\"a}t, Universit{\"a}t Siegen, Siegen, 57068, Germany}
\affil[3]{Ideaded, Carrer de la Tecnologia, 35, Viladecans, Barcelona, 08840, Spain}

\affil[*]{mani.zartab@uab.cat}

\usepackage{url}
\usepackage{hyperref}

\begin{abstract}
For two non-communicating parties, quantum theory can give rise to probability distributions of outcomes that no local classical model can reproduce without communication. However, in the case of two-dimensional systems ($d=2$), it is known that allowing a finite amount of classical communication to shared classical resources makes it possible to simulate these quantum correlations. 
Whether such a simulation remains possible in higher dimensions is still an open question. 
In this work, we identify the key features of the exact classical protocol in $d=2$, and use them to construct robust approximate protocols in higher dimensions.
We assess their performance through a randomized numerical study based on the Total Variation Distance. Our approach exactly reproduces the quantum probability distributions for $d=2$, and performs very well compared to existing protocols for higher dimensions, being the most robust protocol in all cases studied. These results offer new insights into the analytical structure of classical protocols in higher dimensions.
\end{abstract}
\begin{document}

\flushbottom
\maketitle
%
%
\thispagestyle{empty}

\section*{Introduction}

Non-locality, as a fundamental characteristic of nature, is one of the most intriguing consequences of quantum theory \cite{brunner2014bell,scarani2019bell}. Since Bell’s pioneering work, it has been well established that in certain quantum scenarios, two non-communicating, spatially separated parties, say Alice and Bob, can perform independent measurements on shared systems and observe correlations that no local classical model can statistically reproduce \cite{bell1964einstein}. This phenomenon has profound implications for quantum information science, enabling advantages in tasks such as device-independent cryptography \cite{acin2007device} and randomness generation \cite{pironio2010random}. Consequently, understanding and quantifying quantum non-locality is of both foundational and practical importance. 
        
    One approach to quantifying non-locality is through communication complexity, which links quantum correlations to classical communication requirements \cite{brassard2003quantum}. In this framework, the difficulty of classically simulating quantum correlations is measured by the additional amount of information that must be exchanged between two spatially separated parties to reproduce quantum statistics. This perspective not only deepens our understanding of the fundamental limits of classical and quantum theories but also has implications for computational complexity and distributed computing \cite{brassard2003quantum,buhrman2010nonlocality}.
    
    In this context, Toner and Bacon made a significant contribution by proposing a protocol that utilizes only classical resources, \emph{i.e.}, classical shared randomness, plus some finite classical communication. In this case, the shared randomness is randomly distributed three-dimensional real vectors, and a classical communication of one bit, to exactly reproduce the correlations observed in local projective measurements on a shared maximally entangled state—a setting we refer to as the \textit{entanglement scenario} \cite{toner2003communication}. They further demonstrated that in the closely related \textit{prepare-and-measure} (PM) \textit{scenario}, where Alice prepares a state and transmits it to Bob, who then performs a measurement, the quantum statistics of projective measurements on qubits can be reproduced by communicating just two classical bits. More recently, Renner, Tavakoli, and Quintino extended these results to general measurements, \emph{i.e}, Positive Operator-Valued Measures (POVMs), proposing a protocol that maintains the same communication cost \cite{renner2023classical}. A key result of their work is the proof that two bits of communication are both necessary and sufficient for the task.
    
    In higher-dimensional systems $(d \geq 3)$, the challenge of classically simulating quantum correlations becomes more demanding. While exact classical protocols for both the PM scenario and entanglement are well established for qubits $(d = 2)$, extending these results to higher dimensions remains an open problem. In particular, it is still an open question whether classical models, even when augmented with any finite amount of communication, can exactly replicate quantum statistics in these settings.
    
    Nevertheless, motivated by this challenge, several classical models have been proposed that approximate quantum statistics in higher dimensions. For instance, by using a standard sampling tool in statistics, \emph{i.e.,} the rejection method \cite{von1963various,forsythe1972neumann}, we have designed two protocols of communication for simulating the PM scenario, based on two ontological models introduced by Rudolph in  \cite{rudolph2006ontological}, which we refer to as PRUD-1 and PRUD-2. In PRUD-1, the shared randomness consists of rank-1 projectors of dimension $d\geq 3$ weighted similarly to the case in Kochen and Specker's model \cite{kochen2011problem}. In PRUD-2, the problem is embedded in $(d+1)$-dimensional space. Again, the shared randomness is defined via rank-1 projectors, but with a different distribution. Additionally, Montina proposed three protocols in \cite{montina2011approximate}—PMON-1, PMON-2A, and PMON-2B—for the entanglement scenario, extending previous models to higher dimensions. PMON-1 extends Toner and Bacon's protocol 
by reformulating all the steps in terms of pure quantum states, rather than real vectors. PMON-2A and PMON-2B similarly extend Kochen and Specker's model, differing in the nature of shared randomness: PMON-2A employs randomly sampled pure states, while PMON-2B utilizes random bases. Notably, PRUD-1, PMON-1 and PMON-2A are exact for $d=2$. 
    Moreover, protocols designed for the PM scenario can be extended to the entanglement scenario, and vice versa, via Choi–Jamiołkowski isomorphism \cite{adlam2020operational,jiang2013channel,montina2011approximate}.

        In this context, two notable gaps remain in the current literature. First, the protocols developed for the case of $d=2$ \cite{toner2003communication, gisin1999local,degorre2005simulating,renner2023classical,steiner2000towards,cerf2000classical} rely heavily on geometric arguments to show consistency with quantum mechanical predictions, but they do not identify the physical principles that account for their success.
    Secondly, despite offering valuable insights into potential exact strategies, to the best of our knowledge, the performance of the approximate protocols has not been assessed through explicit, randomized numerical studies for both scenarios. Given that these models are inherently approximate for $d\geq3$, it is crucial to evaluate how closely they replicate quantum theory using rigorous distance metrics, such as Kullback–Leibler divergence or Total Variation Distance (TVD). 
    
   This work offers a new perspective on the classical simulations for $d=2$ presented in \cite{toner2003communication,degorre2005simulating,cerf2000classical,renner2023classical,gisin1999local}, and exploits it to construct a probability distribution defined over $SU(d)$ that yields a novel and accurate protocol for simulating projective measurements for $d\geq 2$. Additionally, we conduct a numerical study to compare the performance of our protocol with that of other approximate protocols. Addressing these gaps is essential for advancing our understanding of classical simulations of quantum systems in higher dimensions. 
   
   The numerical assessment is performed for both randomized and structured input setups, considering $d=2,3,4$, and measured by the TVD for both the PM and entanglement scenarios. For the randomized cases, our analysis involves $n=100$ input setups, each with an output sample size of order $N_\textrm{out}\approx 10^{5}$. We show that our protocol exactly reproduces quantum statistics for $d=2$, and consistently ranks among the best-performing models for $d=3$ and $d=4$. In these higher dimensions, it achieves the highest mean accuracy for $d=3$ for both scenarios, and its performance is statistically on par with the other leading protocols, within the statistical error of the mean from the numerical simulations. Moreover, the two additional non-random cases considered, reveal our protocol is the most robust one across all cases, yielding consistently high accuracies.
        
    The structure of this paper is as follows: In the Methods section, we define the PM scenario and entanglement scenario and present our method for generalizing from the exact protocol of $d=2$ to $d\geq3$. The Results section provides numerical results and a comparative analysis of our protocol with existing ones. In the Discussion and Conclusions section, we conclude by discussing key challenges and proposing directions for future research. The detailed numerical reports are reported in the Supplementary Material (SM).

\section*{Methods}
    Quantum mechanics allows for scenarios in which two parties, Alice and Bob, produce statistical correlations that no classical model can reproduce without communication between them \cite{bell1964einstein,clauser1969proposed}.  This raises a fundamental question: if finite communication is permitted, can a classical model reproduce quantum correlations? The amount of communication required can be interpreted as a measure of the advantage that quantum resources offer over classical ones. When such a classical model reproduces quantum correlations exactly using only finite communication, we refer to this as an exact classical simulation.

    Within this context, two closely related quantum scenarios are particularly relevant: the PM scenario and the entanglement scenario. In what follows, we introduce both frameworks and present our classical communication strategy for simulating them.
    \subsection*{PM scenario}
	The PM scenario is defined as follows: Alice prepares a general quantum state of dimension $d$: $ \sigma_\mathrm{A} \in \mathcal{L}(\mathcal{H_A}) $. She sends this state to Bob, who performs a general measurement and reports the outcome $b$, chosen from $m$ possible values. Such measurements are formally described by a POVM, denoted by $\mathcal{B}=\{{B_0,B_1,\cdots,B_{m-1}}\}$, where each operator $B_i$ is positive semi-definite and collectively satisfies the completeness relation: 
    \begin{align}
        \sum_{i=0}^{m-1} B_i\ = I
    \end{align}
\cite{nielsen2010quantum}. If each operator also fulfills the additional constraint $B_i^2 =B_i$, the measurement becomes a projective one, implying the number of outcomes matches the dimension of the Hilbert space, \emph{i.e.}, $m=d$. Independent of the measurement type, the probability of observing outcome $b$ given the quantum state $\sigma_\mathrm{A}$, is determined by Born’s rule: 
    \begin{align}
        \mathrm{P}_{\mathrm{Q}}(b|\sigma_\mathrm{A},\mathcal{B})=\textrm{Tr}[B_b\sigma_\mathrm{A}].
    \end{align}
 
In the classical simulation scenario, we assume that the parties share a hidden random variable $\lambda$, drawn from a uniform probability distribution $\rho(\lambda)$ as shared randomness, and have access to a classical communication channel with finite capacity. The exact shape of  $\lambda$ can vary for different designs. For example, it is three-dimensional real vectors in \cite{toner2003communication,renner2023classical}, random quantum states $|\lambda \rangle$ in \cite{montina2011approximate} , random density operators $|\lambda\rangle \langle \lambda |$ in \cite{rudolph2006ontological}, or random projective measurement basis $\Lambda$ as in our work and \cite{montina2011approximate}. One subset of this arrangement is where each party uses a finite capacity channel only once. In this case, the procedure can be modeled by adding communication to a Local Hidden Variable (LHV) framework as: 
 \begin{align}\label{LHVPM}
     \mathrm{P}_{\mathrm{LHV}}(b|\sigma_\mathrm{A},\mathcal{B})= 
     \int d \lambda \: \rho(\lambda)\: 
     \sum_{c_1=0}^{c_1^{\mathrm{max}}}\sum_{c_2=0}^{c_2^{\mathrm{max}}} \mathrm{P}_{C_1}(c_1|\mathcal{B},\lambda)\mathrm{P}_{C_2} (c_2|c_1, \sigma _A,\lambda) 
      \mathrm{P}_\mathrm{B}(b|c_2,\mathcal{B},\lambda).
 \end{align}
Here $c_1^{\mathrm{max}}$ and $c_2^{\mathrm{max}}$ are the lengths of the messages, while both are assumed to be finite. $\mathrm{P}_{C_1}(c_1|\mathcal{B},\lambda)$ denotes the probability that Bob sends the classical message $c_1$ to Alice, conditioned on the measurement $\mathcal{B}$ and shared randomness $\lambda$. Similarly, $\mathrm{P}_{C_2} (c_2|c_1, \sigma _A,\lambda) $ represents the probability that Alice sends the classical message $c_2$ back to Bob, conditioned on her state $\sigma_\mathrm{A}$, on the message $c_1$, and on the value $\lambda$. Finally, $\mathrm{P}_\mathrm{B}(b|c_2,\mathcal{B},\lambda)$ denotes the probability of Bob obtaining the measurement outcome $b$ conditioned on the measurement $\mathcal{B}$, on the value $\lambda$ of the shared random variable, and the message $c_2$ received from Alice. In our design, $c_1^{\mathrm{max}}=1;c_2^{\mathrm{max}}=d$. Additionally, we let the protocol be occasionally aborted during some runs. This is indicated by setting $c_2=d$ and the output as $b=$\textit{e}. By excluding these runs, we define: 
\begin{align}
   \tilde{\mathrm{P}}_{\mathrm{LHV}}(b|\sigma_\mathrm{A},\mathcal{B}):= \frac{\mathrm{P}_{\mathrm{LHV}}(b|\sigma_\mathrm{A},\mathcal{B})}{1-\mathrm{P}(e|\sigma_\mathrm{A},\mathcal{B})}.
\end{align} 
With this, we state that the model exactly simulates quantum theory if:
\begin{align}
   \tilde{\mathrm{P}}_{\mathrm{LHV}}(b|\sigma_\mathrm{A},\mathcal{B}) =\mathrm{P}_{\mathrm{Q}}(b | \sigma_\mathrm{A} , \mathcal{B}),
\end{align} 
for any arbitrary possible state $\sigma_\mathrm{A}$ and measurement $\mathcal{B}$. For the two-dimensional case, such exact models exist, while for $d\geq 3$, the question remains open.
Approximate strategies, where $\tilde{\mathrm{P}}_{\mathrm{LHV}}(b|\sigma_\mathrm{A},\mathcal{B})\approx \mathrm{P}_{\mathrm{Q}}(b | \sigma_\mathrm{A} , \mathcal{B})$ have been proposed in~\cite{montina2011approximate,rudolph2006ontological}, but systematic studies of their performance have not been carried out. Here, we analyze the exact bi-dimensional models \cite{toner2003communication,renner2023classical,steiner2000towards,degorre2005simulating} from a novel perspective, identify their key ingredients, and devise a protocol capable of simulating quantum statistics for projective measurements and pure states exactly in $d=2$, and with high accuracy in higher dimensions, $d= 3$ and $d=4$.
 
To this end, we first analyze the $d=2$ PRTQ protocol as presented in \cite{renner2023classical} in the case of projective measurements:

    	\begin{itemize}\label{prot-RTQ}
		\item \textbf{Protocol Renner-Tavakoli-Quintino (PRTQ)}

                \textit{Setup}: Alice and Bob are provided, respectively, with the full description of a state represented by the Bloch vector $\vec{x}$, corresponding to the quantum state $\sigma_\mathrm{A}=\frac{I+\vec{x}\cdot \vec{\sigma}}{2}$, and the complete description of a projective measurement characterized by Bloch vectors $\{\vec{y}_b\}$ for the outcomes $b\in \{0,1\}$ ($\vec{y}_0= -\vec{y}_1$) with the vectors normalized as $|\vec{x}|=|\vec{y}_b|=1$. The measurement operators associated with Bob’s measurement are expressed as $B_b = \frac{I+\vec{y}_b\cdot\vec{\sigma}}{2}$. Additionally, both parties have access to two independent sources of randomness, generating real three-dimensional unit vectors $\vec{\lambda}_0$ and $\vec{\lambda}_1$, sampled uniformly from the unit sphere $\mathbb{S}_2$.
            \end{itemize}
        \textit{Protocol:}
		\begin{enumerate}
			\item Bob,
			\begin{enumerate}
				\item Selects one of his measurement vectors $\vec{y}_b$ (for a projective measurement, the choice is arbitrary) 
                \item He sets $c_1 = H(|\vec{y}_b\cdot \vec{\lambda}_0| - |\vec{y}_b\cdot \vec{\lambda}_1|)$, where $H(x)$ is Heaviside step function defined as: $H(x)=1$ if $x\geq 0$ and $H(x)=0$ otherwise;
                \item He sends the classical message $c_1$ to Alice.
			\end{enumerate}
			\item Alice, 
			\begin{enumerate}
				\item receives the message $c_1$, and computes $c_2=H(\vec{x}\cdot\vec{\lambda}_{c_1})$.
                \item She sends the classical bit $c_2$ to Bob
			\end{enumerate} 
			\item Bob,
                \begin{enumerate}
                    \item receives the message $c_2$, and implements the transformation $\vec{\lambda}_{c_1} \mapsto (-1)^{1+c_2}\vec{\lambda}_{c_1} =\vec{\lambda}^\prime$. In other words, he flips the sign of vector $\vec{\lambda}_{c_1}$ if $c_2=0$;
                    \item  Finally, Bob reports the output $b$, with probability
                    \begin{align}
                        \mathrm{P}(b|\vec{\lambda}^\prime,\vec{y}_b)= \frac{ H(\vec{y}_b\cdot \vec{\lambda}^\prime )}{ H(\vec{y}_b\cdot \vec{\lambda}^\prime ) - H(-\vec{y}_b\cdot \vec{\lambda}^\prime)}.
                    \end{align}
                \end{enumerate}
		\end{enumerate}
In step 1, $c_1$ can be regarded as a choice of the shared vector, $\vec{\lambda}_0$ or $\vec{\lambda}_1$, that is more aligned with the direction of $\vec{y}_b$, hence the name \textit{choice method}.

However, an essential implication of this step is that it effectively results in sharing a random vector $\vec{\lambda}$ sampled according to the distribution
\begin{align}\label{eq:2d-dist}
    \rho(\vec{\lambda}|\vec{y}_b)=\mathcal{N} |\vec{\lambda} \cdot \vec{y} _b|.
\end{align}
Here, $\mathcal{N}$ is a normalization factor \cite{degorre2005simulating}. This construction is based on the principles of the rejection method, a standard statistical technique for generating random variables from a target distribution. A detailed explanation is provided in SM, where we demonstrate that it is possible to implement a protocol based on the rejection method in place of the original choice method. The only modification is that Alice and Bob now share a single instance of  $\vec{\lambda}$, and the message $c_{1}$ instead of selecting one random vector, informs Alice whether the current iteration should be accepted or rejected. All other steps of the protocol remain unchanged. It can be proved that the two versions of the protocol are equivalent in terms of both the distribution of outcomes and the average number of shared samples \cite{degorre2005simulating}. Therefore, from now on, we focus the analysis on the shared distribution $\rho(\vec{\lambda}|\vec{y}_b)$ and exploit the rejection method across different protocols to sample from the corresponding target distributions. 

The quantity $|\vec{\lambda} \cdot \vec{y} _b|$, defined as the absolute value of the cosine similarity, serves as a measure of similarity between measurements $\Lambda = \{\Lambda_{\vec{\lambda}},\Lambda_{-\vec{\lambda}}\}$ and $\mathcal{B}=\{B_{\vec{y}_b},B_{-\vec{y}_b}\}$, expressed as:
 \begin{align}\label{cosine-similiarity}
     |\vec{\lambda}\cdot \vec{y}_b| = |\mathrm{P}(\Lambda_{ \vec{\lambda}} | B_{ \vec{y}_b}) - \mathrm{P}(\Lambda_{  - \vec{ \lambda}} | B_{ \vec{y}_b})|,
 \end{align}
where each projector is labeled by its Bloch vector. Here, $\mathrm{P}(\Lambda_{ \vec{\lambda}} | B_{ \vec{y}_b})$ denotes the probability that measuring the state represented by $B_{ \vec{y}_b}$ in the basis $\Lambda$, yields the outcome $\vec{\lambda}$ associated with the projector $\Lambda_{ \vec{\lambda}}$. The magnitude of this probability difference quantifies the measurement correlation: a value of 1 indicates perfect alignment between $\Lambda$ and $\mathcal{B}$, while a value of $0$ signifies complete orthogonality, \emph{i.e.}, that the measurement outcome from one basis provides no information about the other basis. Thus, PRTQ, along with the ones introduced \cite{toner2003communication,degorre2005simulating,steiner2000towards}, uses randomly selected projective measurements $\Lambda$ as a shared random variable, with sampling weights determined by their similarity to the target measurement $\mathcal{B}$. 
For dimension $d=2$, this similarity is uniquely defined by the scalar quantity $|\vec{\lambda}\cdot \vec{y}_b|$.
	
To extend this concept to higher dimensions, consider two random variables $O_\Lambda,O_ \mathcal{B}\in \{0,\dots , d-1\}$ representing the outcome indices of the projective measurements given by $\Lambda=\{\Lambda_i\}_{i=0}^{d-1}$ and $\mathcal{B}=\{B_j\}_{j=0}^{d-1}$, respectively.
Let $\mathrm{P}(O_\Lambda = i|O_\mathcal{B}=j)$ denote the conditional probability of obtaining outcome $i$ under the measurement $O_\Lambda$ given that the outcome $j$ was observed under the measurement $\mathcal{B}$. A natural generalization of the similarity measure in Eq. (\ref{cosine-similiarity}) is given by:
\begin{align}
    \label{eq:similarity}
    D_{\Lambda,\mathcal{B}}(i,j)\coloneqq \mathrm{P}(O_\Lambda = i|O_\mathcal{B}=j) - \mathrm{P}(O_\Lambda \neq i | O_\mathcal{B} = j)
    =2 \mathrm{P}(O_\Lambda = i|O_\mathcal{B}=j) -1.
\end{align}
Thus, the condition that $D_{\Lambda,\mathcal{B}}(i,j) \geq 0$ is equivalent to
\begin{align}\label{halfprobcond}
    \mathrm{P}(O_\Lambda = i|O_\mathcal{B}=j)\geq \frac{1}{2}.
\end{align}

For each outcome $j(i)$, there is at most one outcome $i(j)$ that satisfies this condition for any possible $O_\Lambda$ and $O_\mathcal{B}$. (Observe that only for two-outcome random variables $O_\Lambda$ and $O_\mathcal{B}$ such an outcome always exists.)

A large $  D_{\Lambda,\mathcal{B}}(i,j)$ indicates that whenever the outcome $j$ is observed under $O_\mathcal{B}$, the outcome $i$ under $O_\Lambda$ occurs with high probability. In the limiting cases, where for each $j$, there exist an outcome $i$ such that $D_{\Lambda,\mathcal{B}}(i,j)=1$, the statistics of $O_\Lambda$ and $O_\mathcal{B}$ become effectively equivalent, \emph{i.e.}, knowing the outcome $j\in O_\mathcal{B}$ determines $i\in O_\Lambda$ with absolute certainty. On the other hand, if $D_{\Lambda,\mathcal{B}}(i,j) < 0$ for at least one outcome $j\in O_\mathcal{B}$, this implies that conditioning on $O_\mathcal{B}=j$ does not make any single outcome $i\in O_\Lambda$  more likely than the combined probabilities of all the other outcomes. In such a case, the outcomes of $O_\mathcal{B}$ fail to provide full information about the statistics of $O_\Lambda$.

Before presenting the explicit steps of the protocol, we outline its key components. First, the conditional probabilities are defined, analogously to the two-dimensional case, as  $\mathrm{P}(O_\Lambda = i|O_\mathcal{B}=j) = \textrm{Tr}[\Lambda_i B_j]$. Second, we restrict our attention only to informative events, \emph{i.e.}, those in which the randomly selected shared basis $\Lambda$ satisfies the condition that, for each possible outcome $j$ of $O_{\mathcal{B}}$, there exists at least one outcome $i$ of $O_{{\Lambda}}$ such that $D_{\Lambda,\mathcal{B}}(i,j)\geq 0$. If this condition is not met, the event is rejected. (In the case $d=2$, all events satisfy this condition.)

Third, we must specify a probability distribution shared between Alice and Bob for the accepted events. This probability should depend on the similarity function $D_{\Lambda,\mathcal{B}}(i,j)$ for potentially different pairs $(i,j)$ and it should reduce to a form proportional to $|\vec{\lambda}\cdot \vec{y}_b|$ for $d=2$. A simple functional that satisfies this requirement is  $ D_{\Lambda,\mathcal{B}}(i^*,j^*)^{\alpha_d}$, where $(i^*,j^*)$ denotes the measurement outcomes pair yielding the minimum value of $D_{\Lambda,\mathcal{B}}(i,j)$, and $\alpha_d$ is a parameter that depends on the dimension $d$. For $d=2$, $\alpha_2=1$. For higher dimensions, we observe that from all pairs $(i,j)$, the leading dependence by far is on the term $D_{\Lambda,\mathcal{B}}(i^*,j^*)$, and we find empirically that the optimal exponents in $ D_{\Lambda,\mathcal{B}}(i^*,j^*)^{\alpha_d}$ are $\alpha_3=1/2$ and $\alpha_4=1/4$, suggesting a general scaling of $\alpha_d=2^{2-d}$. Note that as the dimension $d$ increases, the shared distribution becomes increasingly uniform, indicating that the accepted events tend to be more equally informative. 
While more sophisticated variants involving richer dependencies on 
$D_{\Lambda,\mathcal{B}}(i,j)$ values are conceivable and worth exploring, our investigation of various alternative functionals, incorporating contributions of different $(i,j)$ pairs, has shown no improvement in TVD. We obtain that the simple power-law form already captures the key quantum features and yields a protocol that remains robust across different scenarios.

Finally, we have analyzed the role of the threshold parameter 1/2 in the rejection condition $H(\textrm{Tr}[\Lambda_{i^*}B_{j^*}]-\frac{1}{2})$. To this end, We consider a distribution $\rho_d(\Lambda|\mathcal{B})$ proportional to   $(\textrm{Tr}[\Lambda_{i^*}B_{j^*}]-\Delta)^{\alpha_d}H(\textrm{Tr}[\Lambda_{i^*}B_{j^*}]-\Delta)$ with a tunable parameter $\Delta$ close to 1/2. The numerical computation of the TVD reveals that the accuracy reaches its maximum when $\Delta=1/2$, in agreement with our theoretical expectation. The numerical results are provided in Table \ref{threshold} of SM \ref{tables}.

 Therefore, combining the elements discussed above, the shared distribution can be written as
    \begin{align}\label{distribution}
        \rho_d(\Lambda|\mathcal{B})=\: \mathcal{N}_d H(\textrm{Tr}[\Lambda_{i^*}B_{j^*}]-1/2)\times \left(\textrm{Tr}[\Lambda_{i^*}B_{j^*}]-\frac{1}{2}\right)^{\alpha_d},
    \end{align}
where $\mathcal{N}_d$ is the normalization constant that depends on $d$.

We are now ready to present the explicit steps of our protocol.

    	\begin{itemize}\label{P1}
		\item \textbf{P1}
  
            \textit{Setup}: Alice receives the full description of a pure state $\sigma_\mathrm{A} \in \mathcal{L}(\mathcal{H}_\mathrm{A})$ and Bob receives the full description of a projective measurement $\mathcal{B}=\{B_i\}_{i=0}^{d-1}$ and both parties have access to a source that generates the description of a random projective measurement expressed by $\Lambda=\{\Lambda_j\}_{j=0}^{d-1}$. In addition, Bob has access to a random number generator that samples $u\in[0,1]$ uniformly, and sets an appropriate scaling factor $M_d$, required for applying the rejection sampling method (See SM.).
    \end{itemize}
\textit{Protocol:}
        \begin{enumerate}
            \item Bob,
    			\begin{enumerate}
    				\item  Calculates $\textrm{Tr}[\Lambda_{i^*}B_{j^*}]$.
    				\item He sets $c_1=H( \frac{(\textrm{Tr}[\Lambda_{i^*}B_{j^*}]-\frac{1}{2})^{\alpha_d}\times H(\textrm{Tr}[\Lambda_{i^*}B_{j^*}]-\frac{1}{2})} {M_d}-u)$. 
    				\item He sends the bit $c_1$ to Alice.
    			\end{enumerate}
			\item Alice, 
    			\begin{enumerate}
    				\item receives the bit $c_1$. If $c_1=0$, she sets $c_2=d$. Otherwise, she sets $c_2=\arg{\max_{i}\{\textrm{Tr}  [{\Lambda_i\sigma_\mathrm{A}}}]\}$.
                    \item  She sends $c_2$ to Bob.
    			\end{enumerate} 
			\item Bob reports $b = \arg{\max_j\{\textrm{Tr}{[B_j\Lambda_{c_2} ]}\}}$ if $c_2\neq d$. Otherwise, he reports interruption of the protocol by setting $b=e$.
	\end{enumerate}

The operational effect of the protocol can be modeled by:
 \begin{align}\label{P1-Model}
    \tilde{\mathrm{P}}_{LHV}(b|\sigma_\mathrm{A},\mathcal{B})= \int_{SU(d)} d \Lambda \rho_d(\Lambda | \mathcal{B}) 
   \sum_{c_2=0}^{d-1} \mathrm{P}_{C_2}(c_2|\Lambda, \sigma_\mathrm{A}) \mathrm{P}(b|c_2,\mathcal{B} , \Lambda).
 \end{align}  

The initial message $c_1$ serves to share the probability density $\rho_d(\Lambda | \mathcal{B})$ with Alice. The instructions for the following steps are then represented by $\mathrm{P}_{C_2}(c_2|\Lambda, \sigma_\mathrm{A})$ and $\mathrm{P}(b|c_2,\mathcal{B} , \Lambda)$ within the integral.

For $d=2$, one can see that P1 is equivalent to PRTQ with the only difference that the rejection sampling method is used instead of the choice method (See SM.). This becomes apparent by observing that: (\emph{\lroman{1} }) choosing a random vector $\vec{\lambda}$ with $|\vec{\lambda}|=1$ is equivalent to choosing a random projective measurement $\Lambda$. (\emph{\lroman{2} }) Step 2 reduces to simply comparing $u \leq |\cos{\theta}|$ which gives the distribution $\rho (\vec{\lambda}\big| \vec{y}_b)$ proportional to $|\vec{\lambda}.\vec{y}_b|=|\cos{\theta}|$ where $D_{\Lambda,\mathcal{B}}( i^*,j^*)=|\cos{\theta}|$.  (\emph{\lroman{3} }) The condition $D_{\Lambda,\mathcal{B}}( i^*,j^*)\geq \frac{1}{2}$ is always satisfied.  (\emph{\lroman{4} }) In the following steps, sending the message $\arg{\max_{i}\{\textrm{Tr}  [{\Lambda_i\sigma_\mathrm{A}}}]\}\}$ becomes equivalent to Bob inverting the shared vector as  $\vec{\lambda} \mapsto  (-1)^{1+c_2}\vec{\lambda}:=\vec{\lambda}^\prime $ depending on Alice's message, and $\arg{\max_j\{\textrm{Tr}{[B_j\Lambda_{c_2} ]}\}}$ becomes equivalent to selecting the outcome $b$ such that  $H(\vec{\lambda} \cdot \vec{y}_b)=1$, which corresponds to Step 4 of PRTQ.

    \subsection*{Entanglement scenario}
   The entanglement scenario for projective measurements can be expressed as follows: Two parties, each having access to a subsystem of a maximally entangled state $|\Omega\rangle \coloneqq \frac{1}{\sqrt{d}} \sum_i^d | ii \rangle$. They perform a local projective measurement on their subsystems, denoted by $\mathcal{A}$ for Alice and $\mathcal{B}$ for Bob. The joint outcome statistics are governed by the Born rule: $\mathrm{P}_{\mathrm{Q}}(a,b| \mathcal{A}, \mathcal{B}, |\Omega \rangle) = \textrm{Tr}[A_a \otimes  B_b | \Omega \rangle \langle \Omega |]$, where $a$ and $b$ are the outcomes and $A_a\in\mathcal{L}(\mathcal{H}_\mathrm{A})$ and $B_a\in\mathcal{L}(\mathcal{H}_\mathrm{B})$ are the corresponding projectors in $\mathcal{A}$ and $\mathcal{B}$, respectively.

Here, an LHV model augmented with one round of two-way communication can be represented by:
 \begin{align}
     \mathrm{P}_{LHV}(a,b|\mathcal{A},\mathcal{B})= \int \rho(\lambda)   
\sum_{c_1=0}^{c_1^{\mathrm{max}}}\sum_{c_2=0}^{c_2^{\mathrm{max}}} \mathrm{P}_{C_1}(c_1|\mathcal{B},\lambda)\mathrm{P}_{\mathrm{A},C_2}(a,c_2|c_1, \mathcal{A},\lambda) 
      \mathrm{P}_\mathrm{B}(b|c_2,\mathcal{B},\lambda).
 \end{align}
With a slight modification from the model provided in Eq.~(\ref{LHVPM}). In the entanglement scenario, Alice reports an outcome $a$ and sends the classical bit $c_2$ to Bob, with probability $\mathrm{P}_{\mathrm{A},C_2}(a,c_2|c_1, \mathcal{A},\lambda)$. The objective is to construct an LHV model with finite communication such that: 
\begin{align}
   \tilde{\mathrm{P}}_{LHV}(a,b|\mathcal{A},\mathcal{B}):=\frac{\mathrm{P}_{LHV}(a,b|\mathcal{A},\mathcal{B})}{1-\mathrm{P}_{LHV}(e,e|\mathcal{A},\mathcal{B})} 
    =\mathrm{P}_{\mathrm{Q}}(a,b| \mathcal{A}, \mathcal{B}, |\Omega \rangle),
\end{align}for any arbitrary projective measurements $\mathcal{A}$ and $\mathcal{B}$.    
    
This scenario is closely connected to the PM scenario, discussed earlier via Choi–Jamiołkowski isomorphism \cite{adlam2020operational,jiang2013channel}, which establishes a correspondence between positive trace-preserving maps $\mathcal{E}: \mathcal{L}(\mathcal{H}_\mathrm{A}) \rightarrow \mathcal{L}(\mathcal{H}_{\mathrm{B}})$ and a quantum state. Specifically, the Choi matrix associated to $\mathcal{E}$ is defined as $J(\mathcal{E})\equiv \sum_{i,j=0}^{d-1} \mathcal{E}(|i\rangle \langle j |) \otimes |i\rangle \langle j | \in \mathcal{L}(\mathcal{H}_\mathrm{A} \otimes \mathcal{H}_{\mathrm{B}})$ and captures the full action of $\mathcal{E}$, allowing one to translate results between the PM and entanglement scenarios. As a corollary, by applying this isomorphism to the identity map $\mathcal{I}$ acting on $A_a\in \mathcal{L}(\mathcal{H}_\mathrm{A})$, and taking the partial trace over the subspace $\mathrm{B}$, one obtains:
\begin{align}
     \mathcal{I}(A_a)=A_a = \textrm{Tr}_\mathrm{B}[\mathcal{I} \otimes A_a^T J(\mathcal{I})],
 \end{align}
where $J(\mathcal{I})\equiv \sum_{i,j=0}^{d-1} |i\rangle \langle j | \otimes |i\rangle \langle j | \in \mathcal{L}(\mathcal{H}_\mathrm{A} \otimes \mathcal{B}_B)$ is the Choi matrix of the identity map $\mathcal{I}\in \mathcal{L}(\mathcal{H}_\mathrm{A})$, and $A_a^T  \coloneqq \sum_{i,j=0}^{d-1} \langle j |A_a|i\rangle|i\rangle \langle j | \in \mathcal{L}(\mathcal{H}_\mathrm{B})$ is the transpose of $A_a$ in the computational basis. This result implies that any protocol simulating the quantum PM can be converted to one simulating the entanglement scenario, and vice versa.
To establish the link, first notice that: 
	\begin{align}
		\mathrm{P}_{\mathrm{Q}}(a,b| \mathcal{A}, \mathcal{B}, |\Omega \rangle)=
  \mathrm{P}_{\mathrm{Q}}(b|a,\mathcal{A}, \mathcal{B} ,|\Omega\rangle)\mathrm{P}_{\mathrm{Q}}(a|\mathcal{A} , \mathcal{B}, |\Omega\rangle) 
  =\frac{1}{d}\mathrm{P}_{\mathrm{Q}}(b|a,\mathcal{A}, \mathcal{B} ,|\Omega\rangle).
	\end{align}
The second equality follows because the probability of each measurement outcome on subsystem $A$ is independent of subsystem $B$ and is equal to $\frac{1}{d}$. Now, by using the Choi–Jamiołkowski isomorphism:
	\begin{align}\label{Choi}
		\mathrm{P}_{\mathrm{Q}}(b|a,\mathcal{A}, \mathcal{B} ,|\Omega\rangle)= \textrm{Tr}[B_b\textrm{Tr}_\mathrm{A}[\mathcal{I}\otimes A_a J(\mathcal{I})]
  ] 
		=\textrm{Tr}[B_bA_a^T] 
  = \mathrm{P}_{\mathrm{Q}}(b|A_a^T ,\mathcal{B}).
	\end{align}

Therefore, if one considers a projective measurement $\mathcal{A}^T := \{A^T_{a}|a\in \{0,1,\dots,d-1\}\}$, where $A^T_{a=0}=\sigma_\mathrm{A}^T$ and the projectors corresponding to the outcomes $a\neq0$ are arbitrary, the problem of simulating the entanglement scenario becomes linked to the simulation of the PM scenario by conditioning the outcomes on $a=0$ as follows:
\begin{align}
    \textrm{P}_{\mathrm{Q}}&(a=0,b| \mathcal{A}^T, \mathcal{B}, |\Omega \rangle)=\textrm{P}_{\mathrm{Q}}(a=0,b| \mathcal{A}, \mathcal{B}^T, |\Omega \rangle)=\frac{1}{d}\mathrm{P}_{\mathrm{Q}}(b|\sigma_\mathrm{A} ,\mathcal{B}),
\end{align}

where we have used that  $\textrm{Tr}[B_b A^T_a]~=~\textrm{Tr}[B^T_b A_a]$, i.e that one can equivalently consider the measurement $\mathcal{A}$ with $A_{a=0} =\sigma_{\textrm{A}}$, and instead, transpose the elements of $\mathcal{B}$. 
  Now, starting from a protocol originally simulating the entanglement scenario, setting its input to $(\mathcal{A}^T,\mathcal{B})$ or $(\mathcal{A},\mathcal{B}^T)$ and accepting its outputs conditioned on $a=0$, will produce the statistics of the PM scenario for $(\sigma_\textrm{A},\mathcal{B})$. Inversely, if a protocol is originally designed for the PM scenario, the parties run it for the inputs $(A_a^T,\mathcal{B})$ or $(A_a,\mathcal{B}^T)$, where $a$ is selected at each run by rolling an unbiased die of $d$ outcomes. This leads to the simulation of the entanglement scenario for $(\mathcal{A},\mathcal{B})$ \cite{montina2011approximate}.
    With this in hand, we explicitly propose the protocol for simulating the entanglement scenario as follows:

    	\begin{itemize}
    		\item \textbf{P1(entanglement mode)}
    
    \textit{Setup}: Alice and Bob are respectively given the full description of two projective measurements $\mathcal{A}=\{A_i\}_{i=0}^{d-1}$ and $\mathcal{B}=\{B_i\}_{i=0}^{d-1}$; and both also have access to a source of random orthonormal basis generator, producing projective measurements expressed by  $\Lambda=\{\Lambda_j\}_{j=0}^{d-1}$. Additionally, Bob has access to a random number generator that samples $u\in[0,1]$ uniformly, and sets an appropriate scaling factor $M_d$ (See SM.).
    \end{itemize}
\textit{Protocol:}
		\begin{enumerate}
			\item Bob,
			\begin{enumerate}
				\item  calculates $\textrm{Tr}[\Lambda_{i^*}B_{j^*}^T]$.
				\item He sets $c_1=H( \frac{(\textrm{Tr}[\Lambda_{i^*}B_{j^*}^{T}]-\frac{1}{2})^{\alpha_d}\times H(\textrm{Tr}[\Lambda_{i^*}B_{j^*}^{T}]-\frac{1}{2})} {M_d}-u)$.  
				\item He sends the bit $c_1$ to Alice.
			\end{enumerate}
			\item Alice, 
			\begin{enumerate}
				\item receives the bit $c_1$. If $c_1=0$, she sets $c_2=d$ and reports the interruption of the protocol by setting $a=e$. Otherwise, she rolls an unbiased die of $d$ possible outcomes $\{0,\dots,d-1\}$ and reports the output of the die as $a$. Then, she sets $c_2=\arg{\max_{i}\{\textrm{Tr}[{\Lambda_iA_a}}]\}$.
    \item She sends $c_2$ to Bob.
			\end{enumerate} 
			\item Bob reports $b = \arg{\max_j\{\textrm{Tr}{[B^T_j\Lambda_{c_2} ]}\}}$ if $c_2\neq d$. Otherwise, he reports the interruption of the protocol by setting $b=e$.
		\end{enumerate}
    
\section*{Results}

   To compare the performance of the protocols, we conducted a randomized numerical study by running each protocol over a set of $n$ randomly selected pairs of inputs: $\{(\sigma_{A},\mathcal{B})_i\}_{i=1}^n$ for the PM scenario and $\{(\mathcal{A},\mathcal{B})_i\}_{i=1}^n$ for the entanglement scenario. For each setup, the size of the samples of $\Lambda$ is denoted by $N_{\mathrm{ini}}$ and the number of the outputs by $N_{\mathrm{out}}$. For this purpose, we have used the method provided by \cite{ozols2009generate} to generate uniformly distributed $\Lambda$, $\sigma_\mathrm{A}$, $\mathcal{A}$, and $\mathcal{B}$. To assess the performance of the protocols, we use TVD to quantify the distance between the probability distribution predicted by quantum theory and the probability distributions obtained via the classical protocols, which reads
\begin{align}\label{TVD PM scenario}
    \delta=\frac{1}{2n}\sum_{i=1}^n\sum^d_{b=1} |\tilde{\mathrm{P}}_{LHV}(b | \sigma^{i}_\mathrm{A}, \mathcal{B}^{i})-\mathrm{P}_{\mathrm{Q}}(b | \sigma^{i}_\mathrm{A} , \mathcal{B}^{i})|,
\end{align} 
for the PM scenario, and
\begin{align}\label{TVD entanglement scenario}
    \delta=\frac{1}{2n}\sum_{i=1}^n\sum^d_{a,b=1} |\tilde{P}_{LHV}(a,b |\mathcal{A}^{i} , \mathcal{B}^{i})-\mathrm{P}_{\mathrm{Q}}(a,b | \mathcal{A}^{i}, \mathcal{B}^{i})|
\end{align}
for the entanglement scenario.
Note that although the PM and entanglement scenarios are related via the Choi–Jamiołkowski isomorphism, they have fundamentally different sample spaces. Consequently, each protocol has been studied separately in each scenario, as the performance is not necessarily correlated. 

The numerical uncertainty in our estimation of $\delta$ primarily stems from two sources: the finite number of input setups $n$, and the finite sample of output events, $N_{\textrm{out}}$. Furthermore, due to computational limitations, a common set of shared randomness samples was used for all inputs within each protocol. This introduces complex correlations between the TVDs from different input setups, making a rigorous error estimation challenging. With these considerations in mind, we characterize the deviations due to numerical sampling only by the empirical estimate $std_n/\sqrt{n}$ across randomized inputs.

To carry out the runs, we use the rejection method in protocols P1, PRUD-1, and PRUD-2. For P1, we set $M_2=0.5$, and $M_3=M_4=0.7$, and for PRUD-1 and PRUD-2, $M_d=1$ for all $d$'s.  The details of how to choose this parameter are provided in SM.

 \subsection*{d=2}
    
    We have run the protocols 
for $n=100$ uniformly distributed random input setups, with $N_{\mathrm{out}}\approx2.5 \times 10^5$ outputs for both the PM and the entanglement scenarios. The results reveal that, as expected, P1, PMON-1, PMON-2A, and PRUD-1 simulate quantum predictions exactly, with a small deviation of order $\delta \approx 0.001$ that is due to the limited sample size. The exactness of the protocols is further supported by examining the scaling of the $\delta$ with the number of samples, which decreases proportionally to $1/\sqrt{N_{\mathrm{out}}}$, as it should, for any arbitrarily chosen input.

    We also confirm that PMON-2B and PRUD-2 are not exact, which is reflected in the values of the $\delta$. Full results are reported in Table \ref{table-d2} of SM.

\subsection*{d=3}
\begin{figure}[h!]
    \centering
    \includegraphics[width=14cm]{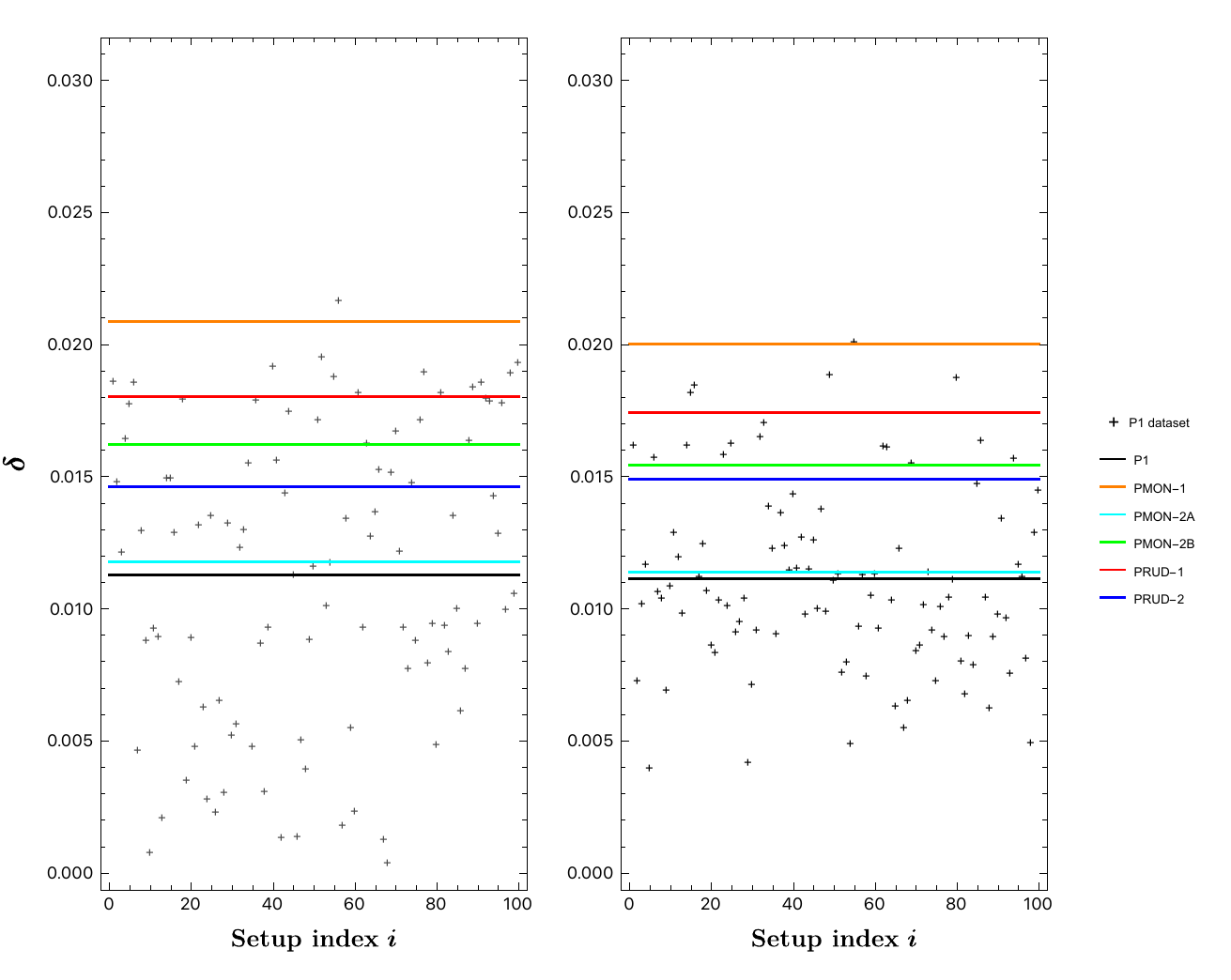}
    \caption{Average performance of the protocols for $d=3$ is shown as lines overlaid on the instances produced by running P1, for $n=100$ randomized inputs—on the left for the PM scenario and on the right for the entanglement scenario. The horizontal axis is the setup index $i\in\{1,\cdots,n\}$ of the input setup $(\sigma_\mathrm{A},\mathcal{B})_i$ for the PM, and $(\mathcal{A},\mathcal{B})_i$ for the entanglement scenario.}
    \label{plot-d3}
\end{figure}

We have analyzed $n=100$ uniformly distributed random input setups, with $N_{\mathrm{out}}\approx 1.2 \times 10^5$ outputs for both the PM and the entanglement scenarios. Our results show that P1 provides the most accurate predictions among all protocols in both scenarios with the lowest $\delta = 0.011$, with PMON-2A following it closely. Figure \ref{plot-d3} summarizes the average performance of all protocols over the randomized inputs studied in this dimension. (Complete numerical data are provided in Table \ref{table-d3} of SM.

\begin{figure}[h!]
    \centering
    \includegraphics[width=12cm]{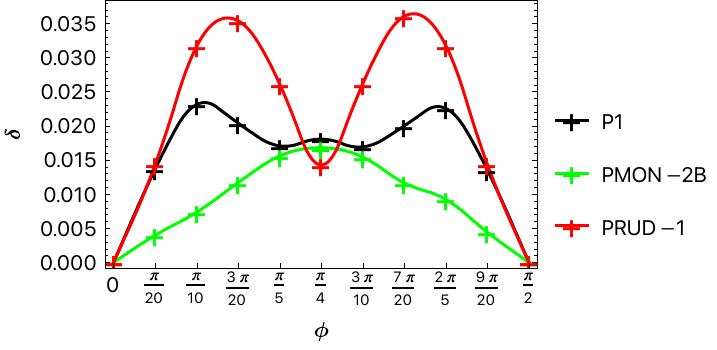}
    \caption{Performance of PMON-2B, PRUD-1, and P1, for the $\phi-$parameterized setup for $d=3$.}
    \label{phi-parameterized-tvd}
\end{figure}

\begin{figure}[h!]
    \centering
		\includegraphics[width=12cm]{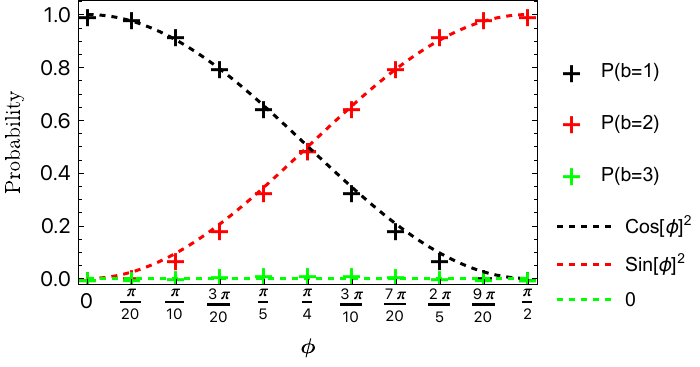}
             \caption{Probability of outcomes of P1 for the $\phi-$parameterized setup for $d=3$, as a function of $\phi$, shown by cross, is compared to the exact values from quantum theory, presented by dashed lines, with $\mathrm{P}_{\mathrm{Q}}(b = 1\big|\sigma_\mathrm{A})=\cos^2{\phi}$, $\mathrm{P}_{\mathrm{Q}}(b = 2\big|\:\sigma_\mathrm{A})=\sin^2{\phi}$, and $\mathrm{P}_{\mathrm{Q}}(b = 3\big|\sigma_\mathrm{A})=0$.}
             \label{phi-parameterized-probabilities}
	\end{figure}

        We also analyze the robustness of the protocols by studying their performance away from random setups and consider structured sources of states and measurements. For the PM scenario, we consider a case where one of the measurement states of $\mathcal{B}$, for example $|b=3\rangle$, has no overlap with the state $|\psi\rangle$ \emph{i.e.}, $|\psi\rangle \in \textrm{span}\{|b=1\rangle, |b=2\rangle\}$ where $\sigma_\mathrm{A} =|\psi\rangle\langle\psi|$. As suggested in \cite{montina2011approximate,rudolph2006ontological}, this structure parameterized solely by $\phi$, makes it convenient to visualize the performance of the protocols as a function~of~$\phi$:
	\begin{align}\label{Parameterization}
		|b = 1\rangle &= \cos{\phi}|\psi \rangle + \sin{\phi} |\psi_{\perp}^{(2)}\rangle  \notag \\
		|b = 2 \rangle &= \sin{\phi}|\psi \rangle - \cos{\phi} |\psi_{\perp}^{(2)}\rangle \notag \\
		|b = 3\rangle &= |\psi_{\perp}^{(3)} \rangle.
	\end{align}
Here $|\psi_{\perp}^{(2)}\rangle$ and  $|\psi_{\perp}^{(3)}\rangle$ are arbitrary perpendicular states, chosen such that $\{\sigma_\mathrm{A},|\psi_{\perp}^{(2)}\rangle\langle \psi_{\perp}^{(2)} |, |\psi_{\perp}^{(3)}\rangle \langle \psi_{\perp}^{(3)}  | \}$ forms a valid projective measurement. We have chosen a set of random $\sigma_\mathrm{A}$'s and then constructed the measurement accordingly. We have tested the performance of the protocols for 11 different values of $\phi$ ($n=11$) in the range $[0,\pi/2]$.
We obtain that PMON-2B and PRUD-2 yield the best values, followed by P1. In Figure \ref{phi-parameterized-tvd}, we have plotted the performance of the best, worst, and our protocol. Additionally, in Figure~\ref{phi-parameterized-probabilities}, we plot the results of P1 compared to the exact quantum predictions. As can be seen, P1 approximates the exact values for the three possible outcomes very well.

	As a structured entanglement scenario, we consider the configuration provided in \cite{collins2002bell}. Alice and Bob each receive two different measurements ($n=4$). The measurements for each party consist of a combination of phase shifts and discrete Fourier transforms, followed by measuring in the computational basis. This setup leads to a maximal violation of the local reality constraint by quantum theory as given by the CGLMP inequalities~\cite{collins2002bell} (a generalization of the CHSH inequalities for $d\geq 3$). We refer to it as the CGLMP setup.
    In this case, the best predictions of the quantum probability distributions come from PRUD-2 and PMON-2B in contrast to the randomized case.
    PMON-1, and P1 follow closely with comparable performance. While the accuracy of P1 remains comparable to the randomized case, the accuracy of PMON-2A, a top performer in that setup, drops significantly here. The increase in its error $\delta$ from the randomized case is approximately double the increase observed for P1, which casts it as the second-worst protocol for this setup. As an additional performance benchmark, we calculated the CGLMP inequality value for each protocol, which can be compared against the quantum~value~2.87. These two particular setups demonstrate a significant change in the order of the best-performing protocols, while P1 maintains robust performance. The detailed numerical report of the simulations for these two special setups are provided in Tables~\ref{phi-parameterized} and~\ref{table-CGLMP} of SM.

\subsection*{d=4}
The performance of all protocols is tested for $n=100$ and $N_{\mathrm{out}} \approx 1 \times 10^5$ randomized inputs, for both scenarios.

\begin{figure}[h!]
    \centering
    \includegraphics[width=14cm]{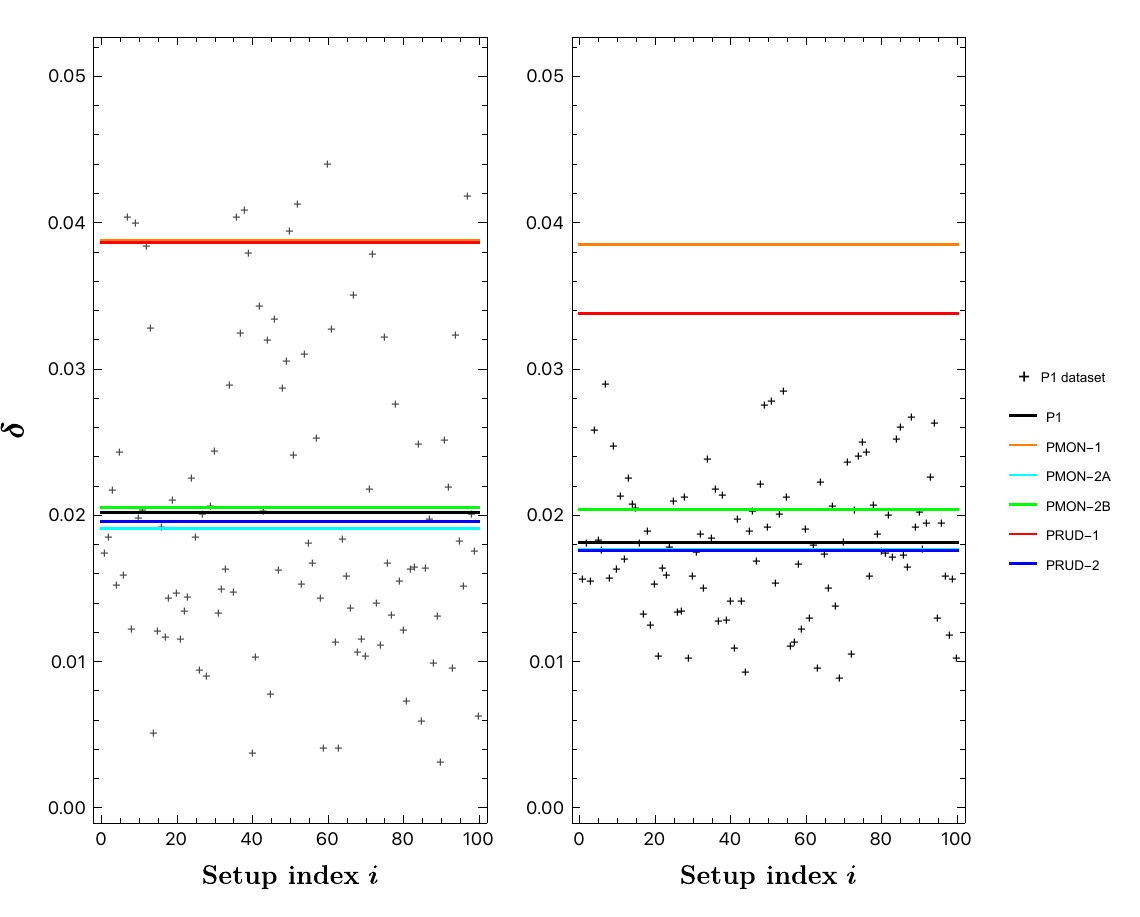}
    \caption{Average performance of the protocols for $d=4$ is shown as lines overlaid on the instances produced by running P1, for $n=100$ randomized inputs—on the left for the PM scenario and on the right for the entanglement scenario. The horizontal axis is the setup index $i\in\{1,\cdots,n\}$ of the input setup $(\sigma_\mathrm{A},\mathcal{B})_i$ for the PM, and $(\mathcal{A},\mathcal{B})_i$ for the entanglement scenario.}
    \label{plot-d4}
\end{figure}

Figure \ref{plot-d4} shows that PMON-2A, PRUD-2, and P1 are the top-performing protocols, with a difference in their performance, comparable to $std_n/\sqrt{n}$. From $d=3$ to $d=4$, while the accuracy of PMON-1 and PRUD-1 degrades significantly, the performances of the other protocols become numerically closer to each other (Compare Tables \ref{table-d3} and  \ref{table-d4} of SM.).

\begin{figure}[h]
    \centering
\includegraphics[width=14cm]{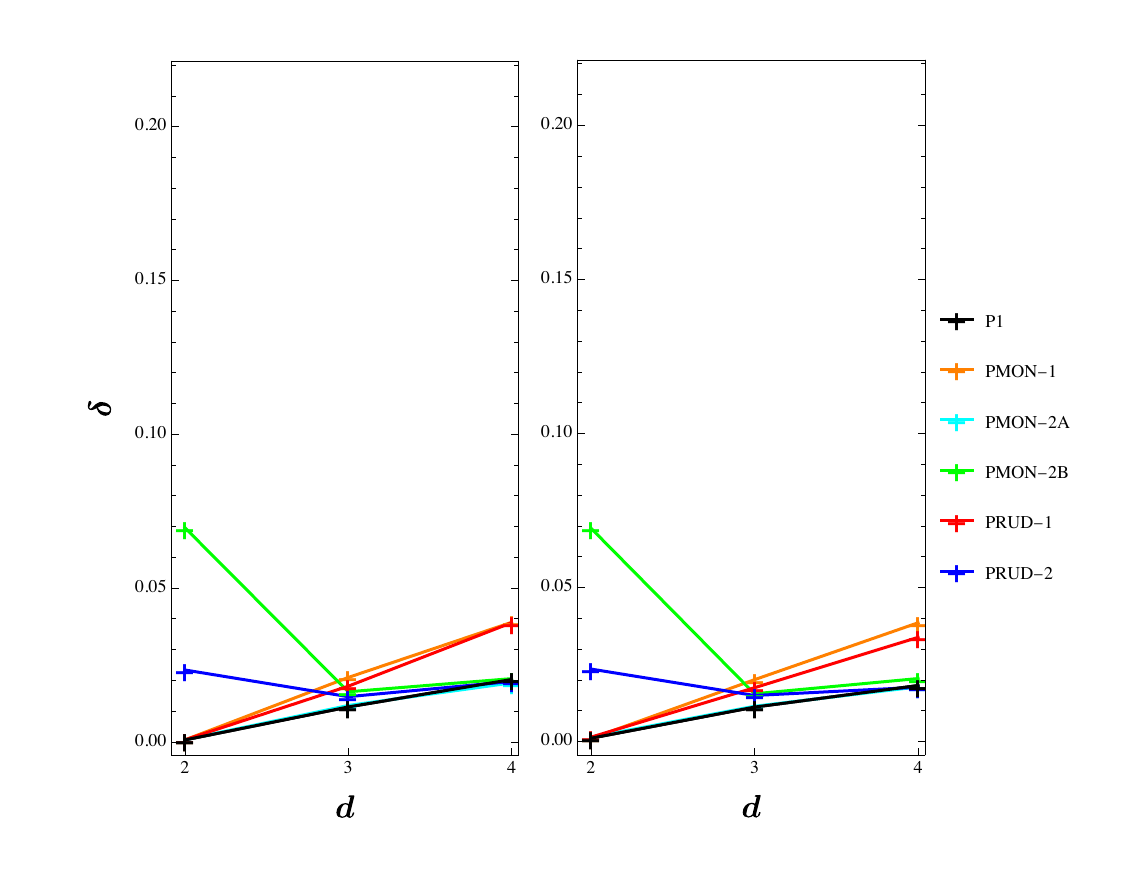}
\caption{Average performance of the protocols as a function of $d$ —on the left for the PM, and on the right for the entanglement scenario.}
\label{plot-dimensional}
\end{figure}

In Figure  \ref{plot-dimensional}, the average performance of the protocols for the randomized input setups is plotted as a function of dimension $d$. It is important to mention $N_{\mathrm{out}}/N_{\mathrm{ini}}$ gets smaller with increasing $d$. As it is shown, protocols P1 and PMON-2A achieve the best performance, on average, while P1, also, turns out to be robust against changing the inputs structure.

These protocols can, in principle, be extended to arbitrary dimensions. However, increasing the dimensionality poses significant challenges in terms of computational resources. Notably, the decrease of the ratio $ N_{\mathrm{out}} / N_{\mathrm{ini}} $ with higher $d$, makes the statistical analysis considerably more demanding. While we expect the trends observed to persist for higher dimensions, a comprehensive numerical investigation remains for future research.

\section*{Discussion and Conclusions}

The exact protocols for qubits introduced in \cite{toner2003communication,degorre2005simulating,cerf2000classical,renner2023classical,gisin1999local} rely heavily on the unique geometric structure of the qubit space, which poses significant challenges when attempting to generalize them to higher dimensions. In this work, we have identified the essential ingredients underlying these protocols and reformulated them in a geometry-independent framework, paving the way for generalization beyond $d=2$.

Our central idea has been to regard the classical simulation process as a weighted sampling over shared randomness. Building on this, we have developed a novel geometry-independent protocol that defines a distribution over $SU(d)$ and provides a new classical communication strategy that closely approximates quantum probability distributions.

To evaluate its performance, we conducted a randomized numerical study comparing our protocol with five existing ones for $d=2,3,4$. Our results show the protocol is exact for $d=2$, and achieves top-tier performance for $d=3$ and $d=4$. It yields the lowest error $\delta$ for $d=3$ among all protocols, and for both $d=3$ and $d=4$, its performance is in line with the other leading protocols, with differences between them being on the order of $std_n/\sqrt{n}$. Furthermore, its ability to maintain high accuracy across all randomized and structured cases shows its superior robustness and better expressivity as a function of $d$. These findings suggest that while an exact protocol for arbitrary $d$ remains elusive, our approximation captures its potential core mechanisms.
%

The similarity function $D_{\Lambda,\mathcal{B}}(i,j)$ defined in Eq.~\eqref{eq:similarity} involves multiple free parameters in $d\geq 3$ (as opposed to $d=2$). To account for this, we explored alternative probability distributions with different contributions of index pairs $(i,j)$ of $D_{\Lambda,\mathcal{B}}(i,j)$ in the probability distribution. However, we have observed no improvement in $\delta$ averaged over randomized inputs. Nevertheless, we expect that additional terms may become relevant as $d$ increases. Analytically, addressing this growing complexity becomes increasingly difficult due to the rapid growth in degrees of freedom. To tackle this, we propose the use of machine learning methods. Indeed, recent advances in deep learning have shown strong potential in identifying classical simulability of quantum distributions \cite{krivachy2020neural} and in constructing local hidden variable models augmented with minimal classical communication \cite{sidajaya2023neural}. Also, our preliminary results in this direction are particularly encouraging: a suitably trained classical neural network, assisted with right set of inputs, appears capable of learning the exact protocol for $d=2$.
The promising role of machine learning hints at hybrid analytical-numerical approaches that could unlock exact protocols in higher dimensions.

\section*{Data availability}
The datasets generated and analyzed during the current study are available in: \hyperref[https://doi.org/10.5281/zenodo.16989149]{\url{https://zenodo.org/records/16989647}}. The code for running the numerical simulations is available in the repository: \hyperref[https://github.com/ManiZart/Prepare-and-measure-and-Entanglement-simulation-beyond-qubits]{\url{https://github.com/ManiZart/Prepare-and-measure-and-Entanglement-simulation-beyond-qubits}}.
\bibliography{References}

\section*{Acknowledgements}

We thank Andreas Winter, Javier Garcia, Niklas Galke, Some Sankar Bhattacharya, Albert Rico, Josep Escrig, and Zuzana Gavorova for their valuable insights and constructive discussions.  We acknowledge financial support from MCIN with funding from the European Union NextGenerationEU (PRTR-C17.I1) and Generalitat de Catalunya, the Ministry of Economic Affairs and Digital Transformation of the Spanish Government through the QUANTUM ENIA project: Quantum Spain, by the European Union through the Recovery, Transformation and Resilience Plan - NextGenerationEU within the framework of the “Digital Spain 2026 Agenda”, and by grant PID2022-141283NB-I00 funded by \mbox{MICIU/AEI/10.13039/501100011033}. MZ acknowledges support from the Ministerio de Ciencia e Innovación of the Spanish Government, contract PRE2020-093634. GG acknowledges DAAD, the Deutsche Forschungsgemeinschaft (DFG, German Research Foundation, project numbers 447948357 and 440958198), the Sino-German Center for Research Promotion (Project M-0294), and the German Ministry of Education and Research (Project QuKuK, BMBF Grant No. 16KIS1618K).

\section*{Author contributions statement}

M.Z. did the main work, G.S. provided the initial ideas, all authors contributed actively in the project, R.M.T. was responsible for the overall supervision. All authors reviewed the manuscript.

\section*{Declarations}

\textbf{Correspondence} and requests for materials should be addressed to M.Z.

\subsection*{Competing interests}
The authors declare no competing interests.

\newpage
\appendix
\section*{Supplementary Material}
\subsection*{Rejection sampling} 
\phantomsection 
\label{Rejection Sampling appendix}
  Rejection sampling \cite{von1963various,forsythe1972neumann} aims to generate samples from a target distribution $\rho(x)$ using a known distribution $\tilde{\rho}(x)$. The standard procedure is as follows:
    \begin{itemize}
    \item \textbf{Rejection sampling}\label{rejectionmethod}
\end{itemize}
\begin{enumerate}
\item draw a number $u\in [0,1]$, uniformly distributed.
\item Generate a sample $x$ from a known probability density function $\tilde{\rho}(x)$ defined over the domain $\mathcal{D}_X$ of the random variable $X$.
\item choose a constant $M \in [1,\infty)$ satisfying $ \rho(x) < M \tilde{\rho}(x)$ for all $ x \in \mathcal{D}_X$.
\item Check whether the condition
\begin{align}
   0<H(\frac{\rho(x)}{M\tilde{\rho}(x)} -u )
\end{align}
 is satisfied:
\begin{itemize}
    \item If yes: accept the sample $x$.
    \item If not: reject the instance and repeat from~1.
\end{itemize}
\end{enumerate}
By doing this procedure, the accepted samples will be distributed by the target distribution $\rho(x)$~. The procedure is stochastic, which means there is a nonzero probability that all samples may be rejected. However, it can be shown that the likelihood of this event decreases exponentially with the number of samples~\cite{steiner2000towards}. 

For the rejection method to work properly, the scaling parameter $M$ should be set correctly. However, to do so, one needs to know the proportion of the normalizations of the initial and target distributions, which in our case is not accessible, especially because of the complicated shape of the target distribution $\rho(\Lambda|\mathcal{B})$. To tackle this challenge, we notice that the shared randomness is initially distributed by the Haar measure $\tilde{\rho}(\Lambda) = \mathcal{N}_{Haar}$. So, we introduce a new parameter $M_d \coloneqq M\mathcal{N}_{Haar}/\mathcal{N}_d$ that casts the comparison step of the rejection sampling algorithm as: 
\begin{align}
    0<H( \frac{(\textrm{Tr}[\Lambda_{i^*}B_{j^*}]-\frac{1}{2})^{\alpha_d}\times H(\textrm{Tr}[\Lambda_{i^*}B_{j^*}]-\frac{1}{2})} {M_d}-u),
\end{align}
which doesn't explicitly depend on the normalization factors $\mathcal{N}_d$ and $\mathcal{N}_{Haar}$. However, we should make sure that $M_d$ is set correctly, which implicitly depends on the normalization factors. To address it, we use the numerical trick that if $M_d$ is set appropriately, multiplying it by a constant $\alpha>1$ will reflect in scaling down the probability of accepting instances by $1/\alpha$. This can be numerically tested by the proportion ${N_{\mathrm{out}}}/{N_{\mathrm{ini}}}$. By scaling $M_d$ by $\alpha = 10$, we have found that our protocol, P1, shows the expected statistics of acceptance ratio if we set  $M_2=0.5$, and $M_3=M_4=0.7$. Besides, for both PRUD-1 and PRUD-2, we have set $M_d=1$ for all $d$'s.

Let us briefly compare the choice method and the rejection method in PRTQ. In the original version, which uses the choice method, $\vec{\lambda}_0$ and  $\vec{\lambda}_1$ are uniformly distributed over $\mathbb{S}^2$. Each $ | \vec{y} _b\cdot\vec{\lambda}_0|$ or $ | \vec{y} _b\cdot\vec{\lambda}_1|$ can take the role of $u$, as they are independently and uniformly distributed over $[0,1]$. Therefore, during each run, one of these quantities effectively serves as the uniform random variable $u$ in the rejection sampling algorithm, while the other plays the role of the target distribution $\rho(\vec{\lambda}|\vec{y}_b)$.  Bob can interchange the roles of $ | \vec{y} _b\cdot\vec{\lambda}_0|$ and  $ | \vec{y} _b\cdot\vec{\lambda}_1|$ ensuring acceptance on every run \cite{degorre2005simulating}. In contrast, the standard rejection method is inherently probabilistic; therefore, the communication 
is only bounded on average. Here, Alice and Bob share only one vector $\vec{\lambda}$, while Bob has an additional local classical resource of a uniformly sampled random number $u\in [0,1]$. In this setup, Bob in step 1 decides whether the sample is accepted or rejected. If rejected, this run of the protocol is aborted. Otherwise, Alice and Bob continue the subsequent steps, and the protocol will produce an outcome. Numerically, these protocols produce the same outcome distributions.
However, in the choice method, all samples are accepted, at the cost of using two independent random classical vectors, while in the rejection method, only one vector is required to be shared at the expense of having rejections. For $d=2$, the probability of acceptance is exactly 1/2  (with $ M_2=0.5$), meaning that the same number of valid outcomes is obtained on average.

\subsection*{Data tables}\label{tables}
The full results of the numerical study are presented in the following tables.
\begin{table}[h!]
\centering

\begin{tabular}{ |c||c | c  | c | c |  }
 \hline
 \multicolumn{5}{|c|}{$d=2$ ; $n=100$} \\
 \hline

 PM scenario &  $N_{\mathrm{ini}}$ & $N_{\mathrm{out}} $ &$  \delta $ & $std_n/\sqrt{n}$\\ 
 &$ (\times 10^{5})$ &$(\times 10^{5})$&$(\times 10^{-2})$ & $(\times 10^{-2})$ \\
 
 \hline

 P1   & 5.0&2.5&   0.056 & 0.004\\
 PMON-1 &     5.0&    2.5& 0.057 & 0.005\\
 PMON-2A  &5.0& 2.5&  0.064 & 0.005\\
 PMON-2B    & 5.0& 2.5& 7.0& 0.3\\
 PRUD-1&     20.0& 2.5& 0.060& 0.005\\
 PRUD-2&   10.0&    2.5& 2.3& 0.1\\
 \hline
  entanglement &  $N_{\mathrm{ini}}$ & $N_{\mathrm{out}} $ &$  \delta $ & $std_n/\sqrt{n}$\\ 
scenario &$ (\times 10^{5})$ &$(\times 10^{5})$&$(\times 10^{-2})$ & $(\times 10^{-2})$ \\
 \hline

 P1   & 5.0&2.5&   0.101 & 0.005\\
 PMON-1 & 2.5&  2.5& 0.092 & 0.005\\
 PMON-2A  & 2.5&  2.5&  0.133& 0.007\\
 PMON-2B    & 2.5&  2.5&  6.9&  0.3\\
 PRUD-1&   20.0 & 2.5& 0.134& 0.006\\
 PRUD-2& 10.0& 2.5& 2.3 & 0.1 \\
 \hline
\end{tabular}
\caption{Performance of the protocols for $n=100$ randomized input setups, for $d=2$, for the PM and entanglement scenarios}
\label{table-d2}
\end{table}
As shown in Table \ref{table-d2}, for $d=2$, P1, PMON-1, PMON-2A, and PRUD-1 result in accurate predictions as proved theoretically. The ratio $N_{\mathrm{out}}/ N_{\mathrm{ini}}<1$ is a sign of stochasticity for the protocols. Notice that as P1 becomes PRTQ with the rejection method of sampling, the fraction $N_{\mathrm{out}}/N_{\mathrm{ini}}=1/2$ reflects the probability of acceptance.
\begin{table}[h!]
\centering

\begin{tabular}{| c||c  |c   |c    | c |}
 \hline
 \multicolumn{5}{|c|}{$d=3$ ; $n=20$} \\
 \hline

 $\Delta$&  $N_{\mathrm{ini}}$ & $N_{\mathrm{out}} $ &$  \delta $&$std_n/\sqrt{n}$  \\ 
 &$ (\times 10^{5})$ &$(\times 10^{4})$&$(\times 10^{-2})$ &$(\times 10^{-2})$ \\
 
 \hline

 $10/24 $& 1.5& 5.9&  5.6  & 0.3\\
 $11/24 $& 2.2 &  6.2 &2.9 & 0.1\\
 $1/2 $&  3.0&  6.0&1.1 &0.1 \\
 $13/24 $&  4.2&  6.1&2.2&0.3\\
 $14/24 $&  6.0&  6.1&4.2  &0.5\\
  \hline
\multicolumn{5}{|c|}{$d=4$ ; $n=20$ } \\
 \hline
  $\Delta$&  $N_{\mathrm{ini}}$ & $N_{\mathrm{out}} $ &$  \delta $ & $std_n/\sqrt{n}$ \\ 
&$ (\times 10^{5})$ &$(\times 10^{4})$&$(\times 10^{-2})$ & $(\times 10^{-2})$ \\
 \hline
 $10/24$& 2.4 & 6.2&6.5 & 0.3\\
 $11/24$& 5.0 & 6.3& 3.1   & 0.2 \\
 $1/2$& 9.0 & 6.1&  2.1& 0.2\\
 $13/24$& 16 & 6.0& 3.3 & 0.5\\
 $14/24$& 30 & 5.9&  5.4 & 0.7\\
 \hline
\end{tabular}
\caption{Performance of P1, for $n=20$ randomized input setups as a function of the cutoff $\Delta$ around $\Delta=\frac{1}{2}$, for $d=3$ and $d=4$}
\label{threshold}
\end{table}

In Table \ref{threshold}, the accuracy of the protocol, as a function of cutoff $\Delta$, is provided. It is evident that by deviating from $1/2$, the accuracy deteriorates significantly. This provides strong evidence that the exclusion cutoff $\Delta$ should indeed be set to 1/2.

\begin{table}[h!]
\centering

\begin{tabular}{ |c||c | c  | c | c |  }
 \hline
 \multicolumn{5}{|c|}{$d=3$ ; $n=100$} \\
 \hline

 PM scenario &  $N_{\mathrm{ini}}$ & $N_{\mathrm{out}} $ &$  \delta $ & $std_n/\sqrt{n}$\\ 
 &$ (\times 10^{5})$ &$(\times 10^{5})$&$(\times 10^{-2})$ & $(\times 10^{-2})$ \\
 
 \hline

 P1   & 6.0&1.2&    1.13&0.06 \\
 PMON-1 &     3.6&    1.2&  2.09 &0.09\\
 PMON-2A  &3.6& 1.2&  1.18 &  0.06\\
 PMON-2B    & 3.6& 1.2&  1.62 & 0.07 \\
 PRUD-1&     36& 1.5&  1.80 & 0.10\\
 PRUD-2&   9.0&    1.1&  1.46 &  0.05\\
 \hline
  entanglement &  $N_{\mathrm{ini}}$ & $N_{\mathrm{out}} $ &$  \delta $ & $std_n/\sqrt{n}$\\ 
scenario &$ (\times 10^{5})$ &$(\times 10^{5})$&$(\times 10^{-2})$ & $(\times 10^{-2})$ \\
 \hline

 P1   & 6.0&1.2&   1.11& 0.03\\
 PMON-1 & 1.2&  1.2&2.00& 0.04\\
 PMON-2A  & 1.2&  1.2&  1.14& 0.04\\
 PMON-2B    & 1.2&  1.2&  1.54 &  0.03\\
 PRUD-1&   36&1.5&1.74& 0.07\\
 PRUD-2& 9.0& 1.1&1.49& 0.03\\
 \hline
\end{tabular}
\caption{Performance of the protocols for $n=100$ randomized input setups, for $d=3$, for the PM and entanglement scenarios}
\label{table-d3}
\end{table}

Table \ref{table-d3} depicts results for the randomized inputs of $d=3$, for both scenarios. As seen, P1 achieves the best performance in both cases, with the lowest attained TVD $\delta$.

The $\phi$-parameterized and the CGLMP setups, as the non-random cases, are reported in Tables \ref{phi-parameterized} and~\ref{table-CGLMP}, respectively. A notable observation is the significant shift in the relative performance of the protocols. In the former case, PMON-2B achieves the lowest error, followed by PRUD-2 and P1, while in contrast, under the randomized setup, PMON-2B ranks as the fourth most accurate protocol. 

The 
 CGLMP inequality is a generalization of Bell-type inequalities to higher-dimensional systems and its maximal quantum violation  is often used to benchmark nonlocality in higher dimensions. 
 For $d=3$, it reads
\begin{align}
\label{eq:CGLMP}
    I_3&\coloneqq \mathrm{P}(\mathcal{A}_1=\mathcal{B}_1) +  \mathrm{P}(\mathcal{A}_2=\mathcal{B}_1-1) + \mathrm{P}(\mathcal{A}_2=\mathcal{B}_2) + \mathrm{P}(\mathcal{A}_1=\mathcal{B}_2) \\ &- \mathrm{P}(\mathcal{A}_1=\mathcal{B}_1 -1) -  \mathrm{P}(\mathcal{A}_2=\mathcal{B}_1) + \mathrm{P}(\mathcal{A}_2=\mathcal{B}_2-1) + \mathrm{P}(\mathcal{A}_1=\mathcal{B}_2+1) \leq 2. \notag 
\end{align}
Inequality~\eqref{eq:CGLMP} is satisfied for any LHV theory, but can be violated by quantum mechanics, with a maximal value of $I_3=2.87$. Here, $\mathrm{P}(\mathcal{A}_i = \mathcal{B}_j+k)\coloneqq \sum_{k=0}^2 \mathrm{P}(a,b+k \textrm{ mod } 3|\mathcal{A}_i , \mathcal{B}_j)$\cite{collins2002bell}. We have numerically simulated the behavior of the protocols for the maximum violation input setup. We see that PMON-2A gives the second-worst accuracy, in comparison to the other protocols, with a noticeably larger $\delta$ compared to P1, despite its performance in the randomized case. Additionally, we computed the CGLMP value for each protocol and compared it to quantum value 2.87, as an additional testbed for the numerical performance of the protocols.
       \begin{table}[h!]
\centering

\begin{tabular}{| c||c  |c   |c  |}
 \hline
 \multicolumn{4}{|c|}{$\phi$-parameterized setup; $d=3$; $n=11$} \\
 \hline  

PM scenario&  $N_{\mathrm{ini}} $ & $N_{\mathrm{out}} $&$  \delta$    \\ \noalign{\vskip -0.2em}
   & $ (\times 10^{6})$&$(\times 10^{5})$& $(\times 10^{-2})$ 
\\
 \hline

 P1   &  3.0& 6.0&    1.50 \\
 PMON-1 &      1.8 &   6.0& 1.72\\
 PMON-2A&1.8&  6.0&  1.55\\
 PMON-2B    &  1.8& 6.0&  0.87\\
 PRUD-1&    18&  7.5& 2.09\\
 PRUD-2&   4.5&    5.6 & 1.10\\
 \hline
\end{tabular}
\caption{Performance of the protocols for the $\phi-$parameterized setup with 11 different values of $\phi\in [0, \frac{\pi}{2}]$, for $d=3$.}
\label{phi-parameterized}
\end{table}

\begin{table}[h!]
\centering
\begin{tabular}{| c||c  |c   |c  |c   |}
 \hline
 \multicolumn{5}{|c|}{CGLMP setup; $d=3$; $n=4$} \\
 \hline  

entanglement
&  $N_{\mathrm{ini}} $ & $N_{\mathrm{out}} $&$  \delta$  & CGLMP  \\ \noalign{\vskip -0.2em}
 scenario  & $ (\times 10^{6})$ &$(\times 10^{5})$& $(\times 10^{-2})$ & value
\\
 \hline

 P1&  3.0&6.0&    1.40 & 2.98 \\
 PMON-1 &     0.6&    6.0 &  1.24 & 2.82\\
 PMON-2A& 0.6&  6.0&   1.67& 2.99\\
 PMON-2B    & 0.6& 6.0&  0.58 & 2.85 \\
 PRUD-1&    18&  7.5&  3.37 &  3.10 \\
 PRUD-2&   4.5&    5.6&0.57& 2.86\\
 \hline
\end{tabular}
\caption{Performance of the protocols for the CGLMP setup, for $d=3$ \cite{collins2002bell}.}
\label{table-CGLMP}
\end{table}

Table \ref{table-d4} reports the values for $d=4$. P1, PMON-2A and PRUD-2, show the highest accuracy with differences of order $\Delta\delta \approx 0.001$, with an overlapping region of $std_n/\sqrt{n} \approx 0.001$. The increase in the $\delta$ from $d=3$ to $d=4$ is noticeable among all protocols.
 \begin{table}[h!]
\centering
\begin{tabular}{ |c||c | c  | c | c |  }
 \hline
 \multicolumn{5}{|c|}{$d=4$ ; $n=100$} \\
 \hline

 PM scenario &  $N_{\mathrm{ini}}$ & $N_{\mathrm{out}} $ &$  \delta $ & $std_n/\sqrt{n}$\\ 
 &$ (\times 10^{6})$ &$ (\times 10^{5})$ &$(\times 10^{-2})$ & $(\times 10^{-2})$ \\
 
 \hline

 P1   & 1.5 & 1.0&  2.02  &   0.10 \\
 PMON-1 &     0.4&     1.0 &  3.88&0.12 \\
 PMON-2A  &0.4&  1.0&   1.91&  0.08\\
 PMON-2B    &  0.4&   1.0& 2.05 &  0.08 \\
 PRUD-1&      7.2&  1.1 &  3.87 &  0.21 \\
 PRUD-2&   1.5&     0.9 &  1.96 &  0.10\\
 \hline
  entanglement &  $N_{\mathrm{ini}}$ & $N_{\mathrm{out}} $ &$  \delta $ & $std_n/\sqrt{n}$\\ 
scenario &$ (\times 10^{6})$ &$ (\times 10^{5})$ &$(\times 10^{-2})$ & $(\times 10^{-2})$ \\
 \hline

 P1   & 1.5& 1.0&   1.81 & 0.05\\
 PMON-1 &  0.1&1.0& 3.85 & 0.04\\
 PMON-2A  & 0.1& 1.0&  1.77 & 0.04 \\
 PMON-2B    &  0.1&   1.0&  2.04 &   0.03\\
 PRUD-1&    7.2&  1.1 & 3.37 &  0.11\\
 PRUD-2&  1.5&  0.9&1.76 &  0.05\\
 \hline
\end{tabular}
\caption{Performance of the protocols for $n=100$ randomized input setups, for $d=4$, for the PM and entanglement scenarios}
\label{table-d4}
\end{table}

\end{document}